\documentclass[twocolumn,aps,prl,superscriptaddress,nofootinbib,showpacs]{revtex4-1}
\usepackage{graphicx}
\usepackage{epstopdf}
\usepackage{color}
\graphicspath{{graphs/}{fig/}}
\usepackage{amsmath}
\usepackage{bm}
\usepackage{slashed}
\usepackage{epsfig}
\usepackage{amsfonts}
\usepackage{epstopdf}
\usepackage{hyperref}
\usepackage{bbm}
\usepackage{textcomp}
\usepackage{multirow}
\usepackage{float}
\usepackage{extarrows}

\newcommand{\sect}[1]{{\it \textbf{#1.} --- }}
\newcommand{\mi}{\mathrm{i}}
\newcommand{\md}{\mathrm{d}}

\newcommand{\cD}{{\mathcal{D}}}
\newcommand{\tD}{{\widetilde{\mathcal{D}}}}
\newcommand{\vtJ}{\overrightarrow{{\hspace{-1mm}\widetilde J}}\hspace{-1mm}}

\begin{document}
\title{Determining Feynman integrals with only input from linear algebra}

\author{Zhi-Feng Liu}
\email{xiangshui@pku.edu.cn}
\affiliation{School of Physics, Peking University, Beijing 100871, China}
\author{Yan-Qing Ma}
\email{yqma@pku.edu.cn}
\affiliation{School of Physics, Peking University, Beijing 100871, China}
\affiliation{Center for High Energy Physics, Peking University, Beijing 100871, China}

\date{\today}
\begin{abstract}
We find that all Feynman integrals (FIs), having any number of loops, can be completely determined once linear relations between FIs are provided. Therefore, FIs computation is conceptually changed to a linear algebraic problem. Examples up to 5 loops are given to verify this observation. As a byproduct, we get a powerful method to calculate perturbative corrections in quantum field theory.
\end{abstract}

\maketitle
\allowdisplaybreaks


\sect{Introduction}
Feynman integrals (FIs) encode  key information of quantum field theories. Study of FIs is important both for exploring mysteries of quantum field theories and for phenomenological application of them. Integrating over some variables is found to be a necessary step to determine FIs in all known systematic methods. This seems to be a reasonable phenomenon, as  FIs themselves are defined by integrating over loop momenta. However, because it is usually hard to perform integration in a systematic and efficient way, is it possible to totally bypass integration in determining FIs?

Systematic methods to compute FIs on the market can be divided into direct methods and indirect methods. Direct methods include sector decomposition~\cite{Hepp:1966eg, Roth:1996pd, Binoth:2000ps, Heinrich:2008si, Smirnov:2015mct, Borowka:2015mxa, Borowka:2017idc,Boos:1990rg, Smirnov:1999gc, Tausk:1999vh, Czakon:2005rk, Smirnov:2009up, Gluza:2007rt},  Mellin-Barnes representation~\cite{Boos:1990rg, Smirnov:1999gc, Tausk:1999vh, Czakon:2005rk, Smirnov:2009up, Gluza:2007rt}, loop-tree duality~\cite{Catani:2008xa,Rodrigo:2008fp,Bierenbaum:2010cy,Bierenbaum:2012th,Tomboulis:2017rvd,Runkel:2019yrs,Capatti:2019ypt,Aguilera-Verdugo:2020set,Song:2021vru,Dubovyk:2022frj}, and so on,  where one computes FIs  by directly performing integration over some variables. Indirect methods compute FIs indirectly by solving corresponding equations, which include
difference equations~\cite{Laporta:2000dsw, Lee:2009dh,Lee:2012te,Lee:2015eva} and differential equations~\cite{Kotikov:1990kg, Kotikov:1991pm, Remiddi:1997ny, Gehrmann:1999as, Argeri:2007up, MullerStach:2012mp, Henn:2013pwa, Henn:2014qga, Moriello:2019yhu, Hidding:2020ytt,Liu:2017jxz,Liu:2020kpc,Liu:2021wks}. To uniquely determine the solution, boundary information are needed in these indirect methods. Unfortunately, the only known systematic way to obtain boundary information is to use direct methods to calculate them. Therefore, integration is still necessary in these indirect methods.

The auxiliary mass flow (AMF) method~\cite{Liu:2017jxz,Liu:2020kpc,Liu:2021wks} is a kind of differential equations method, which computes FIs by setting up and solving differential equations with respect to an auxiliary mass term $\eta$ (called $\eta$-DEs).  The virtue of AMF is that its boundary conditions at $\eta\to\infty$ are simply vacuum bubble integrals, which can be more easily calculated by using other methods~\cite{Davydychev:1992mt, Broadhurst:1998rz, Schroder:2005va, Luthe:2015ngq, Kniehl:2017ikj, Luthe:2017ttc,Baikov:2010hf,Lee:2011jt,Georgoudis:2018olj,Georgoudis:2021onj}.

The observation in this Letter is following. Boundary information for AMF, which can always be casted to single-mass vacuum FIs, can be related to propagator integrals (p-integrals) with one less loops. Then, p-integrals can again be calculated by using the AMF method, with input of new boundary information having one less loops. By using this strategy iteratively, we eventually do not need any input for boundary information in the AMF framework. It is thus surprising to find that integration is totally bypassed in determining FIs.

As a result of our observation, FIs can be completely determined once linear relations between FIs are provided, which are used to decompose all FIs to a small set of bases, called master integrals (MIs), and to set up $\eta$-DEs of these MIs. We note that numerically solving ordinary differential equations (like $\eta$-DEs)  is a well solved mathematical problem \cite{Wason:1987aa}. Therefore, the problem of integrating over loop momenta is now conceptually changed to an linear algebraic problem of exploring the linear space of FIs.

In the rest of the Letter, we first review the AMF method and emphasize its input. We then describe our method to compute boundary conditions within the AMF framework, without any unknown information. Some examples are in order to verify this method. Finally,  we propose a powerful way to calculate perturbative corrections within dimensional regularization.

Before continuing, let us first give a brief introduction to FIs.
A family of FIs  are defined by the following integrals with various values of $\vec{\nu}$,
\begin{align}\label{eq:family}
{I}_{\vec{\nu}}=\int\left(\prod_{i=1}^{L}\frac{\md^{D}\ell_i}{\mi\pi^{D/2}}\right)
\frac{\cD_{K+1}^{-\nu_{K+1}}\cdots \cD_N^{-\nu_N}}
{\cD_1^{\nu_{1}}\cdots \cD_K^{\nu_{K}}},
\end{align}
where $L$ is the number of loops, $\ell_i$ are loop momenta, $D$ is the dimensionality of $\ell_i$, $\cD_1,\ldots,\cD_K$ are inverse propagators with $\nu_1, \ldots, \nu_K$ being integers, and
$\cD_{K+1},\ldots,\cD_N$ are irreducible scalar products introduced for completeness with $\nu_{K+1}, \ldots, \nu_{N}$ being nonpositive integers. It was proved that a family of FIs form a finite-dimensional linear space~\cite{Smirnov:2010hn}. That is, any FI in a give family can be decomposed into a linear combination of MIs, which is a finite set of bases of the linear space formed by the family of FIs. Coefficients in this decomposition are rational functions of all natural variables, like $D$, Mandelstam variables, masses, and the $\eta$ introduced in AMF. Information of the linear space are completely encoded in these decompositions, or linear relations between FIs. Decomposition of FIs is usually realized by integration-by-parts (IBP) reduction, which have been extensively studied ~\cite{Chetyrkin:1981qh, Laporta:2000dsw, Gluza:2010ws, Schabinger:2011dz, vonManteuffel:2012np, Lee:2013mka, vonManteuffel:2014ixa, Larsen:2015ped, Peraro:2016wsq, Mastrolia:2018uzb, Liu:2018dmc,  Guan:2019bcx, Klappert:2019emp, Peraro:2019svx, Frellesvig:2019kgj, Wang:2019mnn, Smirnov:2019qkx, Klappert:2020nbg, Boehm:2020ijp, Basat:2021xnn, Heller:2021qkz, Bendle:2021ueg}.
Having IBP reduction relations, we then only need to study MIs.

Furthermore, because FIs containing linear propagators can be determined by FIs containing only quadratic propagators \cite{Liu:2022tji}, we will not consider  linear propagators anymore.


\sect{The auxiliary mass flow method}
To determine $I_{\vec{\nu}}$ defined in Eq.~\eqref{eq:family},  in the AMF method one introduces an auxiliary family of integrals defined by
\begin{align}
\widetilde{I}_{\vec{\nu}}(\eta)=\int\left(\prod_{i=1}^{L}\frac{\md^{D}\ell_i}{\mi\pi^{D/2}}\right)
\frac{\tD_{K+1}^{-\nu_{K+1}}\cdots \tD_N^{-\nu_N}}
{\tD_1^{\nu_{1}}\cdots \tD_K^{\nu_{K}}}.
\end{align}
Without loss of generality, we assume $\nu_1>0$ and $\cD_1=\ell_1^2-m^2+\mi 0^+$ where $m$ can be zero. We can then choose the propagator mode \cite{Liu:2021wks} to set $\tD_i=\cD_i$ for $i>1$ and modify the mass term for $i=1$ by
\begin{align}
\tD_1=\ell_1^2-m^2- \eta.
\end{align}
Original $I_{\vec{\nu}}$  can be obtained by taking $\eta\to\mi0^-$,
\begin{align}
{I}_{\vec{\nu}}=\lim_{\eta\to \mi0^-} \widetilde{I}_{\vec{\nu}}(\eta) .
\end{align}

Let us denote MIs of the auxiliary family by $\vtJ(\eta)$, and denote its dimension by $n$.
Using IBP reduction, $\frac{\partial}{\partial \eta}\vtJ(\eta)$ can be again expressed as linear combinations of $\vtJ(\eta)$, which results in a system of closed $\eta$-DEs,
\begin{align}\label{eq:deq}
 \frac{\partial}{\partial \eta}\vtJ(\eta) = A(\eta)\vtJ(\eta),
\end{align}
where $A(\eta)$ is an $n\times n$ matrix with entries rationally depending on $\eta$.
Supposing that we already have boundary conditions in hand, we can solve the $\eta$-DEs numerically \cite{Wason:1987aa,Liu:2017jxz} to obtain $\vtJ(\eta)$ and thus their limit $\vtJ(\mi0^-)$. As $\widetilde{I}_{\vec{\nu}}(\eta)$ can be expressed as linear combinations of $\vtJ(\eta)$ using IBP reduction, all original FIs ${I}_{\vec{\nu}}$ (and certainly also their MIs) are eventually determined.

An advantage of AMF is that boundary conditions at $\eta\to \infty$ can be systematically  calculated. In this limit, nonzero contributions only come from integration regions where  linear combinations of loop momenta are either of $\mathcal{O}(\sqrt{|\eta|})$ or $\mathcal{O}(1)$~\cite{Beneke:1997zp,Smirnov:1999bza}. In each of these limited number of regions, a general  propagator can be expressed as 
\begin{align}
\frac{1}{(\ell_\text{L}+\ell_\text{S}+p)^2-m^2-\kappa\,\eta},\nonumber
\end{align}
where $\ell_\text{L}$ is the $\mathcal{O}(\sqrt{\eta})$ part of loop momenta, $\ell_\text{S}$ is the $\mathcal{O}(1)$ part of loop momenta, $p$ is a linear combination of external momenta, $m$ is the mass, and $\kappa=0$ or $1$. Then, if $\ell_\text{L}\neq 0$ or $\kappa\neq 0$, we can simplify the propagator by
\begin{align}\label{eq:mixed1}
\frac{1}{(\ell_\text{L}+\ell_\text{S}+p)^2-m^2-\kappa\,\eta}&\sim \frac{1}{\ell_\text{L}^2-\kappa\,\eta}.
\end{align}
Otherwise, the propagator is unchanged. After the above simplification, 
the resulted new FIs at boundary are either single-mass vacuum FIs or simpler FIs comparing with the original FIs. For the later cases, we can compute them again using AMF, which needs even simpler FIs as input for boundary conditions.

\begin{figure}[htb]
	\centerline{
		\includegraphics[width=0.8\linewidth]{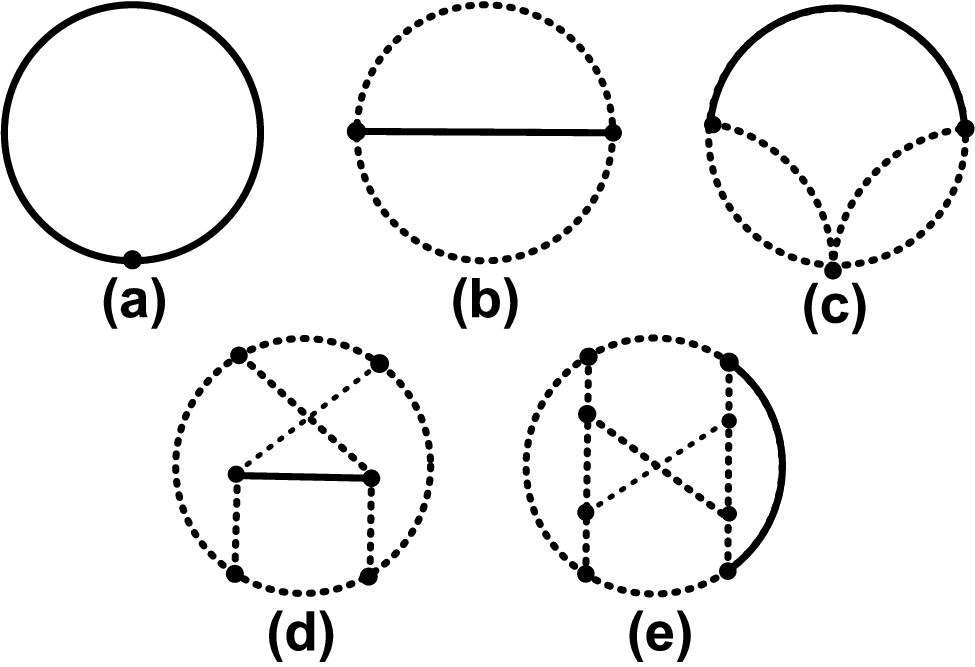}
	}
	\caption{Some typical Feynman diagrams of single-mass vacuum FIs up to 5 loops, where solid lines denote massive propagators, dotted lines denote massless propagators.}\label{fig:bubbles}
\end{figure}

By using AMF iteratively, to determine any $L$-loop FI, we eventually only need single-mass vacuum FIs no more than $L$ loops as additional input besides IBP reductions. Diagrams of some typical single-mass vacuum FIs are shown in Fig.~\ref{fig:bubbles}.

\sect{Determine single-mass vacuum Feynman integrals}
Now let us assume that $I_{\vec{\nu}}$ defined in Eq.~\eqref{eq:family} are single-mass vacuum FIs, with $\cD_1=\ell_1^2-m^2+\mi 0^+$ as the only massive propagator and $\nu_1>0$. Without loss of generality, we set $m^2=1$ in the rest of this Letter.

Let us define a massless p-integral
\begin{align}
	\widehat{I}_{\vec{\nu}^{'}}(\ell_1^2)&=\int\left(\prod_{i=2}^{L}\frac{\md^{D}\ell_i}{\mi\pi^{D/2}}\right)
	\frac{\cD_{K+1}^{-\nu_{K+1}}\cdots \cD_N^{-\nu_N}}
	{\cD_2^{\nu_{2}}\cdots \cD_K^{\nu_{K}}}
\end{align}
with ${\vec{\nu}^{'}}=(\nu_2,\cdots,\nu_N)$, where $\ell_1$ presents as its ``external momentum" and
$\ell_1^2$ is its only mass scale. Based on dimensional counting, we have
\begin{align}\label{eq:scale}
\widehat{I}_{\vec{\nu}^{'}}(\ell_1^2)=
(-\ell_1^2)^{\frac{(L-1)D}{2}-\nu+\nu_1}\widehat{I}_{\vec{\nu}^{'}}(-1),
\end{align}
where $\nu=\sum_{i=1}^N\nu_i$.
The original integral ${I}_{\vec{\nu}}$ is then factorized to two parts and can be evaluated as
\begin{align}\label{eq:drop}
{I}_{\vec{\nu}}=&\int\frac{\md^D\ell_{1}}{\mi\pi^{D/2}}
\frac{(-\ell_1^2)^{\frac{(L-1)D}{2}-\nu+\nu_1}}{(\ell_1^2-1+\mi 0^+)^{\nu_1}} \widehat{I}_{\vec{\nu}^{'}}(-1)\nonumber\\
=&\frac{\Gamma(\nu-LD/2)\Gamma(LD/2-\nu+\nu_1)}
{(-1)^{\nu_1}\Gamma(\nu_1)\Gamma(D/2)} \widehat{I}_{\vec{\nu}^{'}}(-1),
\end{align}
which determines a $L$-loop single-mass vacuum FI ${I}_{\vec{\nu}}$ by a $(L-1)$-loop massless p-integral $\widehat{I}_{\vec{\nu}^{'}}(-1)$. This relation is well-known.

Here comes the key observation: the $(L-1)$-loop massless p-integral $\widehat{I}_{\vec{\nu}^{'}}(-1)$ can be computed via AMF discussed in the last section, which requires single-mass vacuum FIs no more than $(L-1)$ loops as additional input besides IBP reductions. Therefore, we find that, with linear algebra provided by IBP reductions, single-mass vacuum FIs with $L$ loops are determined by that with less than $L$ loops. This works iteratively until  the boundary at $L=1$. Vacuum FIs with $L=1$ are completely determined by the relation~\eqref{eq:drop} by noticing that the value of $0$-loop p-integral is simply $1$.

We eventually arrive at a surprising conclusion that all single-mass vacuum FIs, and therefore all FIs, can be completely determined once linear algebraic relations between different FIs are provided. This conclusion is valid for any number of loops $L$ and arbitrary dimensionality $D$.

\sect{Examples}
To better understand the above observation, let us compute some FIs.

One of the simplest examples is the 2-loop single-mass vacuum integral shown in Fig.~\ref{fig:bubbles} (b), defined by
\begin{equation}
I_{(1,1,1)}=\int\left(\prod_{i=1}^{2}\frac{\md^{D}\ell_i}{\mi\pi^{D/2}}\right)
\frac{1}{(\ell_1^2-1)\ell_2^2(\ell_1+\ell_2)^2},
\end{equation}
where Feynman prescription $\mi 0^+$ for each denominator is suppressed.
The relation \eqref{eq:drop} gives
\begin{align}\label{eq:drop1}
I_{(1,1,1)}=\frac{\Gamma(3-D)\Gamma(D-2)}{-\Gamma(1)\Gamma(D/2)}\widehat{I}_{(1,1)}(-1),
\end{align}
with
\begin{align}
\widehat{I}_{(1,1)}(-1)&=\int\frac{\md^{D}\ell_2}{\mi\pi^{D/2}}
\frac{1}{\ell_2^2(\ell_2+p)^2},
\end{align}
where $p^\mu$ satisfies $p^2=-1$.

To calculate the $1$-loop p-integral $\widehat{I}_{(1,1)}(-1)$ via the AMF method, we introduce auxiliary integrals
\begin{align}
\widetilde{I}_{(1,0)}(\eta)&=\int\frac{\md^{D}\ell_2}{\mi\pi^{D/2}}
\frac{1}{\ell_2^2-\eta},\\
\widetilde{I}_{(1,1)}(\eta)&=\int\frac{\md^{D}\ell_2}{\mi\pi^{D/2}}
\frac{1}{(\ell_2^2-\eta)(\ell_2+p)^2},
\end{align}
which are MIs of the corresponding auxiliary family. Denoting $\vtJ=(\widetilde{I}_{(1,0)},\widetilde{I}_{(1,1)})^T$, $\eta$-DEs can be obtained using IBP reductions,
\begin{align}\label{eq:de1}
	\frac{\partial}{\partial \eta}\vtJ(\eta) = \left(
	\begin{array}{cc}
		\frac{1-\epsilon}{\eta} & 0 \\
		\frac{1-\epsilon}{-\eta(1+\eta)} & \frac{1-2\epsilon}{1+\eta}
	\end{array}
	\right){\vtJ}(\eta).
\end{align}
As  $\eta\to\infty$, only the integration region $|\ell_2|\sim \mathcal{O}(\sqrt{\eta})$ gives nonzero contribution. Thus we have
\begin{align}\label{eq:bc1}
\widetilde{I}_{(1,0)}(\eta)=&{\eta}^{D/2-1}
\int\frac{\md^{D}\ell_2}{\mi\pi^{D/2}}\frac{1}{\ell_2^2-1}
\nonumber\\
=&{\eta}^{D/2-1}
(-1)\Gamma(1-D/2),
\end{align}
where in the last step the relation \eqref{eq:drop} has been used, and
\begin{align}\label{eq:bc2}
\widetilde{I}_{(1,1)}(\eta)
\overset{\eta\to\infty}{=}& \int\frac{\md^{D}\ell_2}{\mi\pi^{D/2}}
\frac{1}{(\ell_2^2-\eta)\ell_2^2}\notag\\
=&{\eta}^{D/2-2}
\int\frac{\md^{D}\ell_2}{\mi\pi^{D/2}}\frac{1}{(\ell_2^2-1)\ell_2^2}
\nonumber\\
=&{\eta}^{D/2-2}
\int\frac{\md^{D}\ell_2}{\mi\pi^{D/2}}\frac{1}{\ell_2^2-1}
\nonumber\\
=&\widetilde{I}_{(1,0)}(\eta) \frac{1}{ \eta},
\end{align}
where scaleless integrals are omitted in the third line.

By solving the $\eta$-DEs \eqref{eq:de1} together with boundary conditions at $\eta\to\infty$ in Eqs. \eqref{eq:bc1} and \eqref{eq:bc2},  $\widehat{I}_{(1,1)}(-1)=\widetilde{I}_{(1,1)}(\mi 0^-)$ is determined. We thus obtain the desired FI $I_{(1,1,1)}$ using the relation~\eqref{eq:drop1}.

Clearly, the same procedure can be used to compute any FI. Let us give the result of another example shown in Fig.~\ref{fig:bubbles} (e), which is one of the most complicated 5-loop single-mass vacuum FIs.
Following the above described procedure, we can compute all MIs in this family to very high precision, with only input of IBP reductions. The result of the corner integral with 10-digit precision is given by
\begin{align}\label{eq:nres}
& - 2.073855510\epsilon^{-2} - 7.812755312\epsilon^{-1}-17.25882864\nonumber\\
&+ 717.6808845 \epsilon+ 8190.876448 \epsilon^2 +78840.29598 \epsilon^3\nonumber\\
&+ 566649.1116 \epsilon^4 + 3901713.802 \epsilon^5+ 23702384.71 \epsilon^6,
\end{align}
where we have set $D = 4-2\epsilon$ with only  9 orders in $\epsilon$ expansion are shown, although more orders and digits can be easily obtained.
The first seven terms of the expansion agree with that obtained in Ref.~\cite{Lee:2011jt}, and other terms are new.

\sect{A new method to calculate perturbative corrections}
An important feature of our strategy is that the FIs we calculate can have arbitrary dimensionality. This on the one hand makes our strategy applicable for a general theory, e.g., nonrelativistic theory with dimensionality equals $3$. And on the other hand, by sampling different dimensionality around a fixed value, say $4-2\epsilon$ with some small values of $\epsilon$, we can fit the Laurent expansion with respect to $\epsilon$ to any desired order, which is actually the way we obtain the results in Eq.~\eqref{eq:nres}. 

If we apply the above strategy directly to physical processes, we arrive at a new and powerful method to calculate perturbative corrections. Let us explain this by an example: the next-to-next-to-leading order (NNLO) QCD correction to top-antitop quark pair fully inclusive production cross section in lepton colliders $e^+e^-\to \gamma^* \to t\bar t+X$, which have been previously calculated in Refs.~\cite{Chetyrkin:1996cf,Gao:2014eea,Chen:2016zbz}. In our method, we calculate bare cross section (before renormalization) with a numerical value of $\epsilon$, and then renormalize it in the standard $\overline{\text{MS}}$ scheme, with the same value of $\epsilon$. To show a numerical result, we choose center-of-mass energy $s$, renormalization scale $\mu$ and top quark mass as $\mu=\sqrt{s}=1$ and $m_t^2=1/8$. We ignore contributions from internal top quark loops and that from  photon interacting with other five type of quarks, because these contributions are very small. Then, if we set $\epsilon=0.001$, the NNLO correction gives
\begin{align}
\sigma_{0.001}^{\text{NNLO}}/(\alpha\alpha_s^2)=&~9.261823090, 
\end{align}
where only 10 digits are shown. Because cross section is a physical quantity that is free of divergence, $\sigma_{\epsilon}^{\text{NNLO}}$ can give an estimation of total cross section up to $\cal{O}(\epsilon)$ error. Now let us calculate the cross section with another value $\epsilon=0.0011$, which gives
\begin{align}
\sigma_{0.0011}^{\text{NNLO}}/(\alpha\alpha_s^2)=&~9.262629688.
\end{align}
The fact that $\sigma_{0.001}^{\text{NNLO}}$ and $\sigma_{0.0011}^{\text{NNLO}}$ has a relative difference at ${\cal{O}}(1/1000)$ level confirms two things. First, $\sigma_{\epsilon}^{\text{NNLO}}$ calculated here is free of $1/\epsilon^n$ divergence, or else the difference should be at ${\cal{O}}(1)$ level. Second, $\sigma_{\epsilon_1}^{\text{NNLO}}=\sigma_{\epsilon_2}^{\text{NNLO}}+{\cal{O}}(\epsilon_1-\epsilon_2)$ is justified. Therefore, we can fit a linear function of $\epsilon$ by combining values of $\sigma_{0.001}^{\text{NNLO}}$ and $\sigma_{0.0011}^{\text{NNLO}}$ to provide a better estimation of $\sigma_{0}^{\text{NNLO}}$,
\begin{align}
\sigma_{0}^{\text{NNLO}}/(\alpha\alpha_s^2)\approx&~ 9.2537+{\cal{O}}(\epsilon^2),
\end{align}
which becomes closer to the exact result $9.253454354$. By calculating each one more value of $\epsilon$, we can further improve the estimation with uncertainty suppressed by one higher order in $\epsilon$.

In this method, we do not need to manipulate a Laurent expansion of $\epsilon$ during the intermediate stage of calculation, and thus the computational time can be usually reduced by several times. This improvement of efficiency is very important for cutting-edge problems. Actually, using this method we have successfully calculated the above mentioned $t\bar{t}$ production to next-to-next-to-next-to-leading order for the first time, which will be presented elsewhere \cite{Chen:2022vzo}.

\sect{Summary and outlook}
By combining the recently proposed AMF method and the relation Eq.~\eqref{eq:drop}, we find that all FIs, with any number of loops and arbitrary dimensionality, can be completely determined once linear relations between FIs are provided. This interesting observation  conceptually changes FIs computation to an algebra problem. This observation has been explicitly verified  by some examples up to 5 loops.

For phenomenological purpose, many general FIs need to be calculated.  The mainstream method on the market to compute FIs can be divided into two steps. In the first step one reduces all FIs to MIs and in the second step one calculates these MIs. Both of the two steps are found to be very difficult for current cutting-edge problems.   With our strategy, IBP reduction becomes the only obstacle for FIs calculation. Our strategy has been implemented in the package \texttt{AMFlow} \cite{Liu:2022chg},  which can fully automatically calculate general FIs, with any number of loops, to high precision, as far as IBP reduction is successful.
These features make our method unique comparing with other methods of FIs computation on the market.

Because FIs with any dimensionality can now be calculated, we can calculate physical processes directly with a given small value of $\epsilon$, the dimensional regulator. In this way, we can significantly improve the efficiency of perturbative calculation. Furthermore, our method is applicable for a general theory, like nonrelativistic theory with dimensionality equals $3$.

\sect{Acknowledgments}
We thank   X. Chen, X. Liu, X. Li, X. Guan and W.H. Wu for many useful communications
and discussions. The work is supported in part by the National Natural Science Foundation of
China (Grants No. 11875071, No. 11975029), the National Key Research and Development Program of China under
Contracts No. 2020YFA0406400, and the High-performance Computing Platform of Peking University.


\providecommand{\href}[2]{#2}\begingroup\raggedright\endgroup


\begin{thebibliography}{10}

\bibitem{Hepp:1966eg}
K.~Hepp, {\it {Proof of the Bogolyubov-Parasiuk theorem on renormalization}},
  \href{http://dx.doi.org/10.1007/BF01773358}{{\em Commun. Math. Phys.}
  {\bfseries 2} (1966) 301--326}
  [\href{http://inspirehep.net/search?p=find+Hepp:1966eg}{{\ttfamily
  InSPIRE}}].

\bibitem{Roth:1996pd}
M.~Roth and A.~Denner, {\it {High-energy approximation of one loop Feynman
  integrals}},  \href{http://dx.doi.org/10.1016/0550-3213(96)00435-X}{{\em
  Nucl. Phys. B} {\bfseries 479} (1996) 495--514}
  [\href{http://arxiv.org/abs/hep-ph/9605420}{{\ttfamily hep-ph/9605420}}]
  [\href{http://inspirehep.net/search?p=find+Roth:1996pd}{{\ttfamily
  InSPIRE}}].

\bibitem{Binoth:2000ps}
T.~Binoth and G.~Heinrich, {\it {An automatized algorithm to compute infrared
  divergent multiloop integrals}},
\href{http://dx.doi.org/10.1016/S0550-3213(00)00429-6}{{\em Nucl. Phys.}
  {\bfseries B585} (2000) 741--759}
  [\href{http://arxiv.org/abs/hep-ph/0004013}{{\ttfamily hep-ph/0004013}}]
  [\href{http://inspirehep.net/search?p=find+Binoth:2000ps}{{\ttfamily
  InSPIRE}}].

\bibitem{Heinrich:2008si}
G.~Heinrich, {\it {Sector Decomposition}},
\href{http://dx.doi.org/10.1142/S0217751X08040263}{{\em Int. J. Mod. Phys.}
  {\bfseries A23} (2008) 1457--1486}
  [\href{http://arxiv.org/abs/0803.4177}{{\ttfamily arXiv:0803.4177}}]
  [\href{http://inspirehep.net/search?p=find+Heinrich:2008si}{{\ttfamily
  InSPIRE}}].

\bibitem{Smirnov:2015mct}
A.~V. Smirnov, {\it {FIESTA4: Optimized Feynman integral calculations with GPU
  support}},
\href{http://dx.doi.org/10.1016/j.cpc.2016.03.013}{{\em Comput. Phys. Commun.}
  {\bfseries 204} (2016) 189--199}
  [\href{http://arxiv.org/abs/1511.03614}{{\ttfamily arXiv:1511.03614}}]
  [\href{http://inspirehep.net/search?p=find+Smirnov:2015mct}{{\ttfamily
  InSPIRE}}].

\bibitem{Borowka:2015mxa}
S.~Borowka, G.~Heinrich, S.~P. Jones, M.~Kerner, J.~Schlenk, and T.~Zirke, {\it
  {SecDec-3.0: numerical evaluation of multi-scale integrals beyond one loop}},
\href{http://dx.doi.org/10.1016/j.cpc.2015.05.022}{{\em Comput. Phys. Commun.}
  {\bfseries 196} (2015) 470--491}
  [\href{http://arxiv.org/abs/1502.06595}{{\ttfamily arXiv:1502.06595}}]
  [\href{http://inspirehep.net/search?p=find+Borowka:2015mxa}{{\ttfamily
  InSPIRE}}].

\bibitem{Borowka:2017idc}
S.~Borowka, G.~Heinrich, S.~Jahn, S.~P. Jones, M.~Kerner, J.~Schlenk, and
  T.~Zirke, {\it {pySecDec: a toolbox for the numerical evaluation of
  multi-scale integrals}},
\href{http://dx.doi.org/10.1016/j.cpc.2017.09.015}{{\em Comput. Phys. Commun.}
  {\bfseries 222} (2018) 313--326}
  [\href{http://arxiv.org/abs/1703.09692}{{\ttfamily arXiv:1703.09692}}]
  [\href{http://inspirehep.net/search?p=find+Borowka:2017idc}{{\ttfamily
  InSPIRE}}].

\bibitem{Boos:1990rg}
E.~E. Boos and A.~I. Davydychev, {\it {A Method of evaluating massive Feynman
  integrals}},  \href{http://dx.doi.org/10.1007/BF01016805}{{\em Theor. Math.
  Phys.} {\bfseries 89} (1991) 1052--1063}
  [\href{http://inspirehep.net/search?p=find+Boos:1990rg}{{\ttfamily
  InSPIRE}}].
[Teor. Mat. Fiz.89,56(1991)].

\bibitem{Smirnov:1999gc}
V.~A. Smirnov, {\it {Analytical result for dimensionally regularized massless
  on shell double box}},
\href{http://dx.doi.org/10.1016/S0370-2693(99)00777-7}{{\em Phys. Lett.}
  {\bfseries B460} (1999) 397--404}
  [\href{http://arxiv.org/abs/hep-ph/9905323}{{\ttfamily hep-ph/9905323}}]
  [\href{http://inspirehep.net/search?p=find+Smirnov:1999gc}{{\ttfamily
  InSPIRE}}].

\bibitem{Tausk:1999vh}
J.~B. Tausk, {\it {Nonplanar massless two loop Feynman diagrams with four
  on-shell legs}},  \href{http://dx.doi.org/10.1016/S0370-2693(99)01277-0}{{\em
  Phys. Lett. B} {\bfseries 469} (1999) 225--234}
  [\href{http://arxiv.org/abs/hep-ph/9909506}{{\ttfamily hep-ph/9909506}}]
  [\href{http://inspirehep.net/search?p=find+Tausk:1999vh}{{\ttfamily
  InSPIRE}}].

\bibitem{Czakon:2005rk}
M.~Czakon, {\it {Automatized analytic continuation of Mellin-Barnes
  integrals}},  \href{http://dx.doi.org/10.1016/j.cpc.2006.07.002}{{\em Comput.
  Phys. Commun.} {\bfseries 175} (2006) 559--571}
  [\href{http://arxiv.org/abs/hep-ph/0511200}{{\ttfamily hep-ph/0511200}}]
  [\href{http://inspirehep.net/search?p=find+Czakon:2005rk}{{\ttfamily
  InSPIRE}}].

\bibitem{Smirnov:2009up}
A.~V. Smirnov and V.~A. Smirnov, {\it {On the Resolution of Singularities of
  Multiple Mellin-Barnes Integrals}},
  \href{http://dx.doi.org/10.1140/epjc/s10052-009-1039-6}{{\em Eur. Phys. J. C}
  {\bfseries 62} (2009) 445--449}
  [\href{http://arxiv.org/abs/0901.0386}{{\ttfamily arXiv:0901.0386}}]
  [\href{http://inspirehep.net/search?p=find+Smirnov:2009up}{{\ttfamily
  InSPIRE}}].

\bibitem{Gluza:2007rt}
J.~Gluza, K.~Kajda, and T.~Riemann, {\it {AMBRE: A Mathematica package for the
  construction of Mellin-Barnes representations for Feynman integrals}},
  \href{http://dx.doi.org/10.1016/j.cpc.2007.07.001}{{\em Comput. Phys.
  Commun.} {\bfseries 177} (2007) 879--893}
  [\href{http://arxiv.org/abs/0704.2423}{{\ttfamily arXiv:0704.2423}}]
  [\href{http://inspirehep.net/search?p=find+Gluza:2007rt}{{\ttfamily
  InSPIRE}}].

\bibitem{Catani:2008xa}
S.~Catani, T.~Gleisberg, F.~Krauss, G.~Rodrigo, and J.-C. Winter, {\it {From
  loops to trees by-passing Feynman's theorem}},
  \href{http://dx.doi.org/10.1088/1126-6708/2008/09/065}{{\em JHEP} {\bfseries
  09} (2008) 065} [\href{http://arxiv.org/abs/0804.3170}{{\ttfamily
  arXiv:0804.3170}}]
  [\href{http://inspirehep.net/search?p=find+Catani:2008xa}{{\ttfamily
  InSPIRE}}].

\bibitem{Rodrigo:2008fp}
G.~Rodrigo, S.~Catani, T.~Gleisberg, F.~Krauss, and J.-C. Winter, {\it {From
  multileg loops to trees (by-passing Feynman's Tree Theorem)}},
  \href{http://dx.doi.org/10.1016/j.nuclphysbps.2008.09.114}{{\em Nucl. Phys. B
  Proc. Suppl.} {\bfseries 183} (2008) 262--267}
  [\href{http://arxiv.org/abs/0807.0531}{{\ttfamily arXiv:0807.0531}}]
  [\href{http://inspirehep.net/search?p=find+Rodrigo:2008fp}{{\ttfamily
  InSPIRE}}].

\bibitem{Bierenbaum:2010cy}
I.~Bierenbaum, S.~Catani, P.~Draggiotis, and G.~Rodrigo, {\it {A Tree-Loop
  Duality Relation at Two Loops and Beyond}},
  \href{http://dx.doi.org/10.1007/JHEP10(2010)073}{{\em JHEP} {\bfseries 10}
  (2010) 073} [\href{http://arxiv.org/abs/1007.0194}{{\ttfamily
  arXiv:1007.0194}}]
  [\href{http://inspirehep.net/search?p=find+Bierenbaum:2010cy}{{\ttfamily
  InSPIRE}}].

\bibitem{Bierenbaum:2012th}
I.~Bierenbaum, S.~Buchta, P.~Draggiotis, I.~Malamos, and G.~Rodrigo, {\it
  {Tree-Loop Duality Relation beyond simple poles}},
  \href{http://dx.doi.org/10.1007/JHEP03(2013)025}{{\em JHEP} {\bfseries 03}
  (2013) 025} [\href{http://arxiv.org/abs/1211.5048}{{\ttfamily
  arXiv:1211.5048}}]
  [\href{http://inspirehep.net/search?p=find+Bierenbaum:2012th}{{\ttfamily
  InSPIRE}}].

\bibitem{Tomboulis:2017rvd}
E.~T. Tomboulis, {\it {Causality and Unitarity via the Tree-Loop Duality
  Relation}},  \href{http://dx.doi.org/10.1007/JHEP05(2017)148}{{\em JHEP}
  {\bfseries 05} (2017) 148} [\href{http://arxiv.org/abs/1701.07052}{{\ttfamily
  arXiv:1701.07052}}]
  [\href{http://inspirehep.net/search?p=find+Tomboulis:2017rvd}{{\ttfamily
  InSPIRE}}].

\bibitem{Runkel:2019yrs}
R.~Runkel, Z.~Sz\H{o}r, J.~P. Vesga, and S.~Weinzierl, {\it {Causality and
  loop-tree duality at higher loops}},
  \href{http://dx.doi.org/10.1103/PhysRevLett.122.111603}{{\em Phys. Rev.
  Lett.} {\bfseries 122} (2019) 111603}
  [\href{http://arxiv.org/abs/1902.02135}{{\ttfamily arXiv:1902.02135}}]
  [\href{http://inspirehep.net/search?p=find+Runkel:2019yrs}{{\ttfamily
  InSPIRE}}]. [Erratum: Phys.Rev.Lett. 123, 059902 (2019)].

\bibitem{Capatti:2019ypt}
Z.~Capatti, V.~Hirschi, D.~Kermanschah, and B.~Ruijl, {\it {Loop-Tree Duality
  for Multiloop Numerical Integration}},
  \href{http://dx.doi.org/10.1103/PhysRevLett.123.151602}{{\em Phys. Rev.
  Lett.} {\bfseries 123} (2019) 151602}
  [\href{http://arxiv.org/abs/1906.06138}{{\ttfamily arXiv:1906.06138}}]
  [\href{http://inspirehep.net/search?p=find+Capatti:2019ypt}{{\ttfamily
  InSPIRE}}].

\bibitem{Aguilera-Verdugo:2020set}
J.~J. Aguilera-Verdugo, F.~Driencourt-Mangin, R.~J. Hern\'andez-Pinto,
  J.~Plenter, S.~Ramirez-Uribe, A.~E. Renteria~Olivo, G.~Rodrigo, G.~F.~R.
  Sborlini, W.~J. Torres~Bobadilla, and S.~Tracz, {\it {Open Loop Amplitudes
  and Causality to All Orders and Powers from the Loop-Tree Duality}},
  \href{http://dx.doi.org/10.1103/PhysRevLett.124.211602}{{\em Phys. Rev.
  Lett.} {\bfseries 124} (2020) 211602}
  [\href{http://arxiv.org/abs/2001.03564}{{\ttfamily arXiv:2001.03564}}]
  [\href{http://inspirehep.net/search?p=find+Aguilera-Verdugo:2020set}{{\ttfamily
  InSPIRE}}].

\bibitem{Song:2021vru}
Q.~Song and A.~Freitas, {\it {On the evaluation of two-loop electroweak box
  diagrams for $e^+e^- \to HZ$ production}},
  \href{http://dx.doi.org/10.1007/JHEP04(2021)179}{{\em JHEP} {\bfseries 04}
  (2021) 179} [\href{http://arxiv.org/abs/2101.00308}{{\ttfamily
  arXiv:2101.00308}}]
  [\href{http://inspirehep.net/search?p=find+Song:2021vru}{{\ttfamily
  InSPIRE}}].

\bibitem{Dubovyk:2022frj}
I.~Dubovyk, A.~Freitas, J.~Gluza, K.~Grzanka, M.~Hidding, and J.~Usovitsch,
  {\it {Evaluation of multi-loop multi-scale Feynman integrals for precision
  physics}},  [\href{http://arxiv.org/abs/2201.02576}{{\ttfamily
  arXiv:2201.02576}}]
  [\href{http://inspirehep.net/search?p=find+Dubovyk:2022frj}{{\ttfamily
  InSPIRE}}].

\bibitem{Laporta:2000dsw}
S.~Laporta, {\it {High precision calculation of multiloop Feynman integrals by
  difference equations}},
  \href{http://dx.doi.org/10.1142/S0217751X00002159}{{\em Int. J. Mod. Phys. A}
  {\bfseries 15} (2000) 5087--5159}
  [\href{http://arxiv.org/abs/hep-ph/0102033}{{\ttfamily hep-ph/0102033}}]
  [\href{http://inspirehep.net/search?p=find+Laporta:2000dsw}{{\ttfamily
  InSPIRE}}].

\bibitem{Lee:2009dh}
R.~N. Lee, {\it {Space-time dimensionality D as complex variable: Calculating
  loop integrals using dimensional recurrence relation and analytical
  properties with respect to D}},
\href{http://dx.doi.org/10.1016/j.nuclphysb.2009.12.025}{{\em Nucl. Phys.}
  {\bfseries B830} (2010) 474--492}
  [\href{http://arxiv.org/abs/0911.0252}{{\ttfamily arXiv:0911.0252}}]
  [\href{http://inspirehep.net/search?p=find+Lee:2009dh}{{\ttfamily InSPIRE}}].

\bibitem{Lee:2012te}
R.~N. Lee and V.~A. Smirnov, {\it {The Dimensional Recurrence and Analyticity
  Method for Multicomponent Master Integrals: Using Unitarity Cuts to Construct
  Homogeneous Solutions}},
  \href{http://dx.doi.org/10.1007/JHEP12(2012)104}{{\em JHEP} {\bfseries 12}
  (2012) 104} [\href{http://arxiv.org/abs/1209.0339}{{\ttfamily
  arXiv:1209.0339}}]
  [\href{http://inspirehep.net/search?p=find+Lee:2012te}{{\ttfamily InSPIRE}}].

\bibitem{Lee:2015eva}
R.~N. Lee and K.~T. Mingulov, {\it {Introducing SummerTime: a package for
  high-precision computation of sums appearing in DRA method}},
  \href{http://dx.doi.org/10.1016/j.cpc.2016.02.018}{{\em Comput. Phys.
  Commun.} {\bfseries 203} (2016) 255--267}
  [\href{http://arxiv.org/abs/1507.04256}{{\ttfamily arXiv:1507.04256}}]
  [\href{http://inspirehep.net/search?p=find+Lee:2015eva}{{\ttfamily
  InSPIRE}}].

\bibitem{Kotikov:1990kg}
A.~V. Kotikov, {\it {Differential equations method: New technique for massive
  Feynman diagrams calculation}},
\href{http://dx.doi.org/10.1016/0370-2693(91)90413-K}{{\em Phys. Lett.}
  {\bfseries B254} (1991) 158--164}
  [\href{http://inspirehep.net/search?p=find+Kotikov:1990kg}{{\ttfamily
  InSPIRE}}].

\bibitem{Kotikov:1991pm}
A.~V. Kotikov, {\it {Differential equation method: The Calculation of N point
  Feynman diagrams}},
  \href{http://dx.doi.org/10.1016/0370-2693(91)90536-Y}{{\em Phys. Lett. B}
  {\bfseries 267} (1991) 123--127}
  [\href{http://inspirehep.net/search?p=find+Kotikov:1991pm}{{\ttfamily
  InSPIRE}}]. [Erratum: Phys.Lett.B 295, 409--409 (1992)].

\bibitem{Remiddi:1997ny}
E.~Remiddi, {\it {Differential equations for Feynman graph amplitudes}},
{\em Nuovo Cim.} {\bfseries A110} (1997) 1435--1452
  [\href{http://arxiv.org/abs/hep-th/9711188}{{\ttfamily hep-th/9711188}}]
  [\href{http://inspirehep.net/search?p=find+Remiddi:1997ny}{{\ttfamily
  InSPIRE}}].

\bibitem{Gehrmann:1999as}
T.~Gehrmann and E.~Remiddi, {\it {Differential equations for two loop four
  point functions}},
\href{http://dx.doi.org/10.1016/S0550-3213(00)00223-6}{{\em Nucl. Phys.}
  {\bfseries B580} (2000) 485--518}
  [\href{http://arxiv.org/abs/hep-ph/9912329}{{\ttfamily hep-ph/9912329}}]
  [\href{http://inspirehep.net/search?p=find+Gehrmann:1999as}{{\ttfamily
  InSPIRE}}].

\bibitem{Argeri:2007up}
M.~Argeri and P.~Mastrolia, {\it {Feynman Diagrams and Differential
  Equations}},  \href{http://dx.doi.org/10.1142/S0217751X07037147}{{\em Int. J.
  Mod. Phys. A} {\bfseries 22} (2007) 4375--4436}
  [\href{http://arxiv.org/abs/0707.4037}{{\ttfamily arXiv:0707.4037}}]
  [\href{http://inspirehep.net/search?p=find+Argeri:2007up}{{\ttfamily
  InSPIRE}}].

\bibitem{MullerStach:2012mp}
S.~M\"uller-Stach, S.~Weinzierl, and R.~Zayadeh, {\it {Picard-Fuchs equations
  for Feynman integrals}},
\href{http://dx.doi.org/10.1007/s00220-013-1838-3}{{\em Commun. Math. Phys.}
  {\bfseries 326} (2014) 237--249}
  [\href{http://arxiv.org/abs/1212.4389}{{\ttfamily arXiv:1212.4389}}]
  [\href{http://inspirehep.net/search?p=find+MullerStach:2012mp}{{\ttfamily
  InSPIRE}}].

\bibitem{Henn:2013pwa}
J.~M. Henn, {\it {Multiloop integrals in dimensional regularization made
  simple}},
\href{http://dx.doi.org/10.1103/PhysRevLett.110.251601}{{\em Phys. Rev. Lett.}
  {\bfseries 110} (2013) 251601}
  [\href{http://arxiv.org/abs/1304.1806}{{\ttfamily arXiv:1304.1806}}]
  [\href{http://inspirehep.net/search?p=find+Henn:2013pwa}{{\ttfamily
  InSPIRE}}].

\bibitem{Henn:2014qga}
J.~M. Henn, {\it {Lectures on differential equations for Feynman integrals}},
\href{http://dx.doi.org/10.1088/1751-8113/48/15/153001}{{\em J. Phys.}
  {\bfseries A48} (2015) 153001}
  [\href{http://arxiv.org/abs/1412.2296}{{\ttfamily arXiv:1412.2296}}]
  [\href{http://inspirehep.net/search?p=find+Henn:2014qga}{{\ttfamily
  InSPIRE}}].

\bibitem{Moriello:2019yhu}
F.~Moriello, {\it {Generalised power series expansions for the elliptic planar
  families of Higgs + jet production at two loops}},
  \href{http://dx.doi.org/10.1007/JHEP01(2020)150}{{\em JHEP} {\bfseries 01}
  (2020) 150} [\href{http://arxiv.org/abs/1907.13234}{{\ttfamily
  arXiv:1907.13234}}]
  [\href{http://inspirehep.net/search?p=find+Moriello:2019yhu}{{\ttfamily
  InSPIRE}}].

\bibitem{Hidding:2020ytt}
M.~Hidding, {\it {DiffExp, a Mathematica package for computing Feynman
  integrals in terms of one-dimensional series expansions}},
  \href{http://dx.doi.org/10.1016/j.cpc.2021.108125}{{\em Comput. Phys.
  Commun.} {\bfseries 269} (2021) 108125}
  [\href{http://arxiv.org/abs/2006.05510}{{\ttfamily arXiv:2006.05510}}]
  [\href{http://inspirehep.net/search?p=find+Hidding:2020ytt}{{\ttfamily
  InSPIRE}}].

\bibitem{Liu:2017jxz}
X.~Liu, Y.-Q. Ma, and C.-Y. Wang, {\it {A Systematic and Efficient Method to
  Compute Multi-loop Master Integrals}},
\href{http://dx.doi.org/10.1016/j.physletb.2018.02.026}{{\em Phys. Lett.}
  {\bfseries B779} (2018) 353--357}
  [\href{http://arxiv.org/abs/1711.09572}{{\ttfamily arXiv:1711.09572}}]
  [\href{http://inspirehep.net/search?p=find+Liu:2017jxz}{{\ttfamily
  InSPIRE}}].

\bibitem{Liu:2020kpc}
X.~Liu, Y.-Q. Ma, W.~Tao, and P.~Zhang, {\it {Calculation of Feynman loop
  integration and phase-space integration via auxiliary mass flow}},
  \href{http://dx.doi.org/10.1088/1674-1137/abc538}{{\em Chin. Phys. C}
  {\bfseries 45} (2021) 013115}
  [\href{http://arxiv.org/abs/2009.07987}{{\ttfamily arXiv:2009.07987}}]
  [\href{http://inspirehep.net/search?p=find+Liu:2020kpc}{{\ttfamily
  InSPIRE}}].

\bibitem{Liu:2021wks}
X.~Liu and Y.-Q. Ma, {\it {Multiloop corrections for collider processes using
  auxiliary mass flow}},
  \href{http://dx.doi.org/10.1103/PhysRevD.105.L051503}{{\em Phys. Rev. D}
  {\bfseries 105} (2022) L051503}
  [\href{http://arxiv.org/abs/2107.01864}{{\ttfamily arXiv:2107.01864}}]
  [\href{http://inspirehep.net/search?p=find+Liu:2021wks}{{\ttfamily
  InSPIRE}}].

\bibitem{Davydychev:1992mt}
A.~I. Davydychev and J.~B. Tausk, {\it {Two loop selfenergy diagrams with
  different masses and the momentum expansion}},
\href{http://dx.doi.org/10.1016/0550-3213(93)90338-P}{{\em Nucl. Phys.}
  {\bfseries B397} (1993) 123--142}
  [\href{http://inspirehep.net/search?p=find+Davydychev:1992mt}{{\ttfamily
  InSPIRE}}].

\bibitem{Broadhurst:1998rz}
D.~J. Broadhurst, {\it {Massive three - loop Feynman diagrams reducible to SC*
  primitives of algebras of the sixth root of unity}},
\href{http://dx.doi.org/10.1007/s100529900935}{{\em Eur. Phys. J.} {\bfseries
  C8} (1999) 311--333} [\href{http://arxiv.org/abs/hep-th/9803091}{{\ttfamily
  hep-th/9803091}}]
  [\href{http://inspirehep.net/search?p=find+Broadhurst:1998rz}{{\ttfamily
  InSPIRE}}].

\bibitem{Schroder:2005va}
Y.~Schr\"{o}der and A.~Vuorinen, {\it {High-precision epsilon expansions of
  single-mass-scale four-loop vacuum bubbles}},
\href{http://dx.doi.org/10.1088/1126-6708/2005/06/051}{{\em JHEP} {\bfseries
  06} (2005) 051} [\href{http://arxiv.org/abs/hep-ph/0503209}{{\ttfamily
  hep-ph/0503209}}]
  [\href{http://inspirehep.net/search?p=find+Schroder:2005va}{{\ttfamily
  InSPIRE}}].

\bibitem{Luthe:2015ngq}
T.~Luthe, {\em {Fully massive vacuum integrals at 5 loops}}.
\newblock PhD thesis, Bielefeld U.,
\newblock 2015
  [\href{http://inspirehep.net/search?p=find+Luthe:2015ngq}{{\ttfamily
  InSPIRE}}].
\newblock
\url{https://pub.uni-bielefeld.de/publication/2776013}.
\newblock

\bibitem{Kniehl:2017ikj}
B.~A. Kniehl, A.~F. Pikelner, and O.~L. Veretin, {\it {Three-loop massive
  tadpoles and polylogarithms through weight six}},
\href{http://dx.doi.org/10.1007/JHEP08(2017)024}{{\em JHEP} {\bfseries 08}
  (2017) 024} [\href{http://arxiv.org/abs/1705.05136}{{\ttfamily
  arXiv:1705.05136}}]
  [\href{http://inspirehep.net/search?p=find+Kniehl:2017ikj}{{\ttfamily
  InSPIRE}}].

\bibitem{Luthe:2017ttc}
T.~Luthe, A.~Maier, P.~Marquard, and Y.~Schroder, {\it {Complete
  renormalization of QCD at five loops}},
\href{http://dx.doi.org/10.1007/JHEP03(2017)020}{{\em JHEP} {\bfseries 03}
  (2017) 020} [\href{http://arxiv.org/abs/1701.07068}{{\ttfamily
  arXiv:1701.07068}}]
  [\href{http://inspirehep.net/search?p=find+Luthe:2017ttc}{{\ttfamily
  InSPIRE}}].

\bibitem{Baikov:2010hf}
P.~A. Baikov and K.~G. Chetyrkin, {\it {Four Loop Massless Propagators: An
  Algebraic Evaluation of All Master Integrals}},
  \href{http://dx.doi.org/10.1016/j.nuclphysb.2010.05.004}{{\em Nucl. Phys. B}
  {\bfseries 837} (2010) 186--220}
  [\href{http://arxiv.org/abs/1004.1153}{{\ttfamily arXiv:1004.1153}}]
  [\href{http://inspirehep.net/search?p=find+Baikov:2010hf}{{\ttfamily
  InSPIRE}}].

\bibitem{Lee:2011jt}
R.~N. Lee, A.~V. Smirnov, and V.~A. Smirnov, {\it {Master Integrals for
  Four-Loop Massless Propagators up to Transcendentality Weight Twelve}},
  \href{http://dx.doi.org/10.1016/j.nuclphysb.2011.11.005}{{\em Nucl. Phys. B}
  {\bfseries 856} (2012) 95--110}
  [\href{http://arxiv.org/abs/1108.0732}{{\ttfamily arXiv:1108.0732}}]
  [\href{http://inspirehep.net/search?p=find+Lee:2011jt}{{\ttfamily InSPIRE}}].

\bibitem{Georgoudis:2018olj}
A.~Georgoudis, V.~Goncalves, E.~Panzer, and R.~Pereira, {\it {Five-loop
  massless propagator integrals}},
  [\href{http://arxiv.org/abs/1802.00803}{{\ttfamily arXiv:1802.00803}}]
  [\href{http://inspirehep.net/search?p=find+Georgoudis:2018olj}{{\ttfamily
  InSPIRE}}].

\bibitem{Georgoudis:2021onj}
A.~Georgoudis, V.~Gon\c{c}alves, E.~Panzer, R.~Pereira, A.~V. Smirnov, and
  V.~A. Smirnov, {\it {Glue-and-cut at five loops}},
  \href{http://dx.doi.org/10.1007/JHEP09(2021)098}{{\em JHEP} {\bfseries 09}
  (2021) 098} [\href{http://arxiv.org/abs/2104.08272}{{\ttfamily
  arXiv:2104.08272}}]
  [\href{http://inspirehep.net/search?p=find+Georgoudis:2021onj}{{\ttfamily
  InSPIRE}}].

\bibitem{Wason:1987aa}
W.~Wason, {\em Asymptotic Expansions for Ordinary Differential Equations}.
\newblock Dover Publications, Inc.,
\newblock 1987
  [\href{http://inspirehep.net/search?p=find+Wason:1987aa}{{\ttfamily
  InSPIRE}}].

\bibitem{Smirnov:2010hn}
A.~V. Smirnov and A.~V. Petukhov, {\it {The Number of Master Integrals is
  Finite}},
\href{http://dx.doi.org/10.1007/s11005-010-0450-0}{{\em Lett. Math. Phys.}
  {\bfseries 97} (2011) 37--44}
  [\href{http://arxiv.org/abs/1004.4199}{{\ttfamily arXiv:1004.4199}}]
  [\href{http://inspirehep.net/search?p=find+Smirnov:2010hn}{{\ttfamily
  InSPIRE}}].

\bibitem{Chetyrkin:1981qh}
K.~G. Chetyrkin and F.~V. Tkachov, {\it {Integration by Parts: The Algorithm to
  Calculate beta Functions in 4 Loops}},
\href{http://dx.doi.org/10.1016/0550-3213(81)90199-1}{{\em Nucl. Phys.}
  {\bfseries B192} (1981) 159--204}
  [\href{http://inspirehep.net/search?p=find+Chetyrkin:1981qh}{{\ttfamily
  InSPIRE}}].

\bibitem{Gluza:2010ws}
J.~Gluza, K.~Kajda, and D.~A. Kosower, {\it {Towards a Basis for Planar
  Two-Loop Integrals}},
\href{http://dx.doi.org/10.1103/PhysRevD.83.045012}{{\em Phys. Rev.} {\bfseries
  D83} (2011) 045012} [\href{http://arxiv.org/abs/1009.0472}{{\ttfamily
  arXiv:1009.0472}}]
  [\href{http://inspirehep.net/search?p=find+Gluza:2010ws}{{\ttfamily
  InSPIRE}}].

\bibitem{Schabinger:2011dz}
R.~M. Schabinger, {\it {A New Algorithm For The Generation Of
  Unitarity-Compatible Integration By Parts Relations}},
\href{http://dx.doi.org/10.1007/JHEP01(2012)077}{{\em JHEP} {\bfseries 01}
  (2012) 077} [\href{http://arxiv.org/abs/1111.4220}{{\ttfamily
  arXiv:1111.4220}}]
  [\href{http://inspirehep.net/search?p=find+Schabinger:2011dz}{{\ttfamily
  InSPIRE}}].

\bibitem{vonManteuffel:2012np}
A.~von Manteuffel and C.~Studerus,
{\it {Reduze 2 - Distributed Feynman Integral Reduction}},
  [\href{http://arxiv.org/abs/1201.4330}{{\ttfamily arXiv:1201.4330}}]
  [\href{http://inspirehep.net/search?p=find+vonManteuffel:2012np}{{\ttfamily
  InSPIRE}}].

\bibitem{Lee:2013mka}
R.~N. Lee, {\it {LiteRed 1.4: a powerful tool for reduction of multiloop
  integrals}},
\href{http://dx.doi.org/10.1088/1742-6596/523/1/012059}{{\em J. Phys. Conf.
  Ser.} {\bfseries 523} (2014) 012059}
  [\href{http://arxiv.org/abs/1310.1145}{{\ttfamily arXiv:1310.1145}}]
  [\href{http://inspirehep.net/search?p=find+Lee:2013mka}{{\ttfamily
  InSPIRE}}].

\bibitem{vonManteuffel:2014ixa}
A.~von Manteuffel and R.~M. Schabinger, {\it {A novel approach to integration
  by parts reduction}},
\href{http://dx.doi.org/10.1016/j.physletb.2015.03.029}{{\em Phys. Lett.}
  {\bfseries B744} (2015) 101--104}
  [\href{http://arxiv.org/abs/1406.4513}{{\ttfamily arXiv:1406.4513}}]
  [\href{http://inspirehep.net/search?p=find+vonManteuffel:2014ixa}{{\ttfamily
  InSPIRE}}].

\bibitem{Larsen:2015ped}
K.~J. Larsen and Y.~Zhang, {\it {Integration-by-parts reductions from unitarity
  cuts and algebraic geometry}},
\href{http://dx.doi.org/10.1103/PhysRevD.93.041701}{{\em Phys. Rev.} {\bfseries
  D93} (2016) 041701} [\href{http://arxiv.org/abs/1511.01071}{{\ttfamily
  arXiv:1511.01071}}]
  [\href{http://inspirehep.net/search?p=find+Larsen:2015ped}{{\ttfamily
  InSPIRE}}].

\bibitem{Peraro:2016wsq}
T.~Peraro, {\it {Scattering amplitudes over finite fields and multivariate
  functional reconstruction}},
\href{http://dx.doi.org/10.1007/JHEP12(2016)030}{{\em JHEP} {\bfseries 12}
  (2016) 030} [\href{http://arxiv.org/abs/1608.01902}{{\ttfamily
  arXiv:1608.01902}}]
  [\href{http://inspirehep.net/search?p=find+Peraro:2016wsq}{{\ttfamily
  InSPIRE}}].

\bibitem{Mastrolia:2018uzb}
P.~Mastrolia and S.~Mizera, {\it {Feynman Integrals and Intersection Theory}},
\href{http://dx.doi.org/10.1007/JHEP02(2019)139}{{\em JHEP} {\bfseries 02}
  (2019) 139} [\href{http://arxiv.org/abs/1810.03818}{{\ttfamily
  arXiv:1810.03818}}]
  [\href{http://inspirehep.net/search?p=find+Mastrolia:2018uzb}{{\ttfamily
  InSPIRE}}].

\bibitem{Liu:2018dmc}
X.~Liu and Y.-Q. Ma, {\it {Determining arbitrary Feynman integrals by vacuum
  integrals}},  \href{http://dx.doi.org/10.1103/PhysRevD.99.071501}{{\em Phys.
  Rev. D} {\bfseries 99} (2019) 071501}
  [\href{http://arxiv.org/abs/1801.10523}{{\ttfamily arXiv:1801.10523}}]
  [\href{http://inspirehep.net/search?p=find+Liu:2018dmc}{{\ttfamily
  InSPIRE}}].

\bibitem{Guan:2019bcx}
X.~Guan, X.~Liu, and Y.-Q. Ma, {\it {Complete reduction of integrals in
  two-loop five-light-parton scattering amplitudes}},
  \href{http://dx.doi.org/10.1088/1674-1137/44/9/093106}{{\em Chin. Phys. C}
  {\bfseries 44} (2020) 093106}
  [\href{http://arxiv.org/abs/1912.09294}{{\ttfamily arXiv:1912.09294}}]
  [\href{http://inspirehep.net/search?p=find+Guan:2019bcx}{{\ttfamily
  InSPIRE}}].

\bibitem{Klappert:2019emp}
J.~Klappert and F.~Lange, {\it {Reconstructing rational functions with
  FireFly}},  \href{http://dx.doi.org/10.1016/j.cpc.2019.106951}{{\em Comput.
  Phys. Commun.} {\bfseries 247} (2020) 106951}
  [\href{http://arxiv.org/abs/1904.00009}{{\ttfamily arXiv:1904.00009}}]
  [\href{http://inspirehep.net/search?p=find+Klappert:2019emp}{{\ttfamily
  InSPIRE}}].

\bibitem{Peraro:2019svx}
T.~Peraro, {\it {FiniteFlow: multivariate functional reconstruction using
  finite fields and dataflow graphs}},
\href{http://dx.doi.org/10.1007/JHEP07(2019)031}{{\em JHEP} {\bfseries 07}
  (2019) 031} [\href{http://arxiv.org/abs/1905.08019}{{\ttfamily
  arXiv:1905.08019}}]
  [\href{http://inspirehep.net/search?p=find+Peraro:2019svx}{{\ttfamily
  InSPIRE}}].

\bibitem{Frellesvig:2019kgj}
H.~Frellesvig, F.~Gasparotto, S.~Laporta, M.~K. Mandal, P.~Mastrolia,
  L.~Mattiazzi, and S.~Mizera, {\it {Decomposition of Feynman Integrals on the
  Maximal Cut by Intersection Numbers}},
\href{http://dx.doi.org/10.1007/JHEP05(2019)153}{{\em JHEP} {\bfseries 05}
  (2019) 153} [\href{http://arxiv.org/abs/1901.11510}{{\ttfamily
  arXiv:1901.11510}}]
  [\href{http://inspirehep.net/search?p=find+Frellesvig:2019kgj}{{\ttfamily
  InSPIRE}}].

\bibitem{Wang:2019mnn}
Y.~Wang, Z.~Li, and N.~Ul~Basat, {\it {Direct reduction of multiloop multiscale
  scattering amplitudes}},
  \href{http://dx.doi.org/10.1103/PhysRevD.101.076023}{{\em Phys. Rev. D}
  {\bfseries 101} (2020) 076023}
  [\href{http://arxiv.org/abs/1901.09390}{{\ttfamily arXiv:1901.09390}}]
  [\href{http://inspirehep.net/search?p=find+Wang:2019mnn}{{\ttfamily
  InSPIRE}}].

\bibitem{Smirnov:2019qkx}
A.~V. Smirnov and F.~S. Chuharev, {\it {FIRE6: Feynman Integral REduction with
  Modular Arithmetic}},
  \href{http://dx.doi.org/10.1016/j.cpc.2019.106877}{{\em Comput. Phys.
  Commun.} {\bfseries 247} (2020) 106877}
  [\href{http://arxiv.org/abs/1901.07808}{{\ttfamily arXiv:1901.07808}}]
  [\href{http://inspirehep.net/search?p=find+Smirnov:2019qkx}{{\ttfamily
  InSPIRE}}].

\bibitem{Klappert:2020nbg}
J.~Klappert, F.~Lange, P.~Maierh\"ofer, and J.~Usovitsch, {\it {Integral
  reduction with Kira 2.0 and finite field methods}},
  \href{http://dx.doi.org/10.1016/j.cpc.2021.108024}{{\em Comput. Phys.
  Commun.} {\bfseries 266} (2021) 108024}
  [\href{http://arxiv.org/abs/2008.06494}{{\ttfamily arXiv:2008.06494}}]
  [\href{http://inspirehep.net/search?p=find+Klappert:2020nbg}{{\ttfamily
  InSPIRE}}].

\bibitem{Boehm:2020ijp}
J.~Boehm, M.~Wittmann, Z.~Wu, Y.~Xu, and Y.~Zhang, {\it {IBP reduction
  coefficients made simple}},
  \href{http://dx.doi.org/10.1007/JHEP12(2020)054}{{\em JHEP} {\bfseries 12}
  (2020) 054} [\href{http://arxiv.org/abs/2008.13194}{{\ttfamily
  arXiv:2008.13194}}]
  [\href{http://inspirehep.net/search?p=find+Boehm:2020ijp}{{\ttfamily
  InSPIRE}}].

\bibitem{Basat:2021xnn}
N.~u. Basat, Z.~Li, and Y.~Wang, {\it {Reduction of the planar double-box
  diagram for single-top production via auxiliary mass flow}},
  \href{http://dx.doi.org/10.1103/PhysRevD.104.056020}{{\em Phys. Rev. D}
  {\bfseries 104} (2021) 056020}
  [\href{http://arxiv.org/abs/2102.08225}{{\ttfamily arXiv:2102.08225}}]
  [\href{http://inspirehep.net/search?p=find+Basat:2021xnn}{{\ttfamily
  InSPIRE}}].

\bibitem{Heller:2021qkz}
M.~Heller and A.~von Manteuffel, {\it {MultivariateApart: Generalized partial
  fractions}},  \href{http://dx.doi.org/10.1016/j.cpc.2021.108174}{{\em Comput.
  Phys. Commun.} {\bfseries 271} (2022) 108174}
  [\href{http://arxiv.org/abs/2101.08283}{{\ttfamily arXiv:2101.08283}}]
  [\href{http://inspirehep.net/search?p=find+Heller:2021qkz}{{\ttfamily
  InSPIRE}}].

\bibitem{Bendle:2021ueg}
D.~Bendle, J.~Boehm, M.~Heymann, R.~Ma, M.~Rahn, L.~Ristau, M.~Wittmann, Z.~Wu,
  and Y.~Zhang, {\it {Two-loop five-point integration-by-parts relations in a
  usable form}},  [\href{http://arxiv.org/abs/2104.06866}{{\ttfamily
  arXiv:2104.06866}}]
  [\href{http://inspirehep.net/search?p=find+Bendle:2021ueg}{{\ttfamily
  InSPIRE}}].

\bibitem{Liu:2022tji}
Z.-F. Liu and Y.-Q. Ma, {\it {Automatic computation of Feynman integrals
  containing linear propagators via auxiliary mass flow}},
  \href{http://dx.doi.org/10.1103/PhysRevD.105.074003}{{\em Phys. Rev. D}
  {\bfseries 105} (2022) 074003}
  [\href{http://arxiv.org/abs/2201.11636}{{\ttfamily arXiv:2201.11636}}]
  [\href{http://inspirehep.net/search?p=find+Liu:2022tji}{{\ttfamily
  InSPIRE}}].

\bibitem{Beneke:1997zp}
M.~Beneke and V.~A. Smirnov, {\it {Asymptotic expansion of Feynman integrals
  near threshold}},
\href{http://dx.doi.org/10.1016/S0550-3213(98)00138-2}{{\em Nucl.Phys.}
  {\bfseries B522} (1998) 321--344}
  [\href{http://arxiv.org/abs/hep-ph/9711391}{{\ttfamily hep-ph/9711391}}]
  [\href{http://inspirehep.net/search?p=find+Beneke:1997zp}{{\ttfamily
  InSPIRE}}].

\bibitem{Smirnov:1999bza}
V.~A. Smirnov, {\it {Problems of the strategy of regions}},
  \href{http://dx.doi.org/10.1016/S0370-2693(99)01061-8}{{\em Phys. Lett. B}
  {\bfseries 465} (1999) 226--234}
  [\href{http://arxiv.org/abs/hep-ph/9907471}{{\ttfamily hep-ph/9907471}}]
  [\href{http://inspirehep.net/search?p=find+Smirnov:1999bza}{{\ttfamily
  InSPIRE}}].

\bibitem{Chetyrkin:1996cf}
K.~G. Chetyrkin, J.~H. Kuhn, and M.~Steinhauser, {\it {Three loop polarization
  function and $O(\alpha_s^2)$ corrections to the production of heavy quarks}},
   \href{http://dx.doi.org/10.1016/S0550-3213(96)00534-2}{{\em Nucl. Phys. B}
  {\bfseries 482} (1996) 213--240}
  [\href{http://arxiv.org/abs/hep-ph/9606230}{{\ttfamily hep-ph/9606230}}]
  [\href{http://inspirehep.net/search?p=find+Chetyrkin:1996cf}{{\ttfamily
  InSPIRE}}].

\bibitem{Gao:2014eea}
J.~Gao and H.~X. Zhu, {\it {Top Quark Forward-Backward Asymmetry in $e^+e^-$
  Annihilation at Next-to-Next-to-Leading Order in QCD}},
  \href{http://dx.doi.org/10.1103/PhysRevLett.113.262001}{{\em Phys. Rev.
  Lett.} {\bfseries 113} (2014) 262001}
  [\href{http://arxiv.org/abs/1410.3165}{{\ttfamily arXiv:1410.3165}}]
  [\href{http://inspirehep.net/search?p=find+Gao:2014eea}{{\ttfamily
  InSPIRE}}].

\bibitem{Chen:2016zbz}
L.~Chen, O.~Dekkers, D.~Heisler, W.~Bernreuther, and Z.-G. Si, {\it {Top-quark
  pair production at next-to-next-to-leading order QCD in electron positron
  collisions}},  \href{http://dx.doi.org/10.1007/JHEP12(2016)098}{{\em JHEP}
  {\bfseries 12} (2016) 098} [\href{http://arxiv.org/abs/1610.07897}{{\ttfamily
  arXiv:1610.07897}}]
  [\href{http://inspirehep.net/search?p=find+Chen:2016zbz}{{\ttfamily
  InSPIRE}}].

\bibitem{Chen:2022vzo}
X.~Chen, X.~Guan, C.-Q. He, X.~Liu, and Y.-Q. Ma, {\it {Heavy-quark-pair
  production at lepton colliders at NNNLO in QCD}},
  [\href{http://arxiv.org/abs/2209.14259}{{\ttfamily arXiv:2209.14259}}]
  [\href{http://inspirehep.net/search?p=find+Chen:2022vzo}{{\ttfamily
  InSPIRE}}].

\bibitem{Liu:2022chg}
X.~Liu and Y.-Q. Ma, {\it {AMFlow: a Mathematica Package for Feynman integrals
  computation via Auxiliary Mass Flow}},
  [\href{http://arxiv.org/abs/2201.11669}{{\ttfamily arXiv:2201.11669}}]
  [\href{http://inspirehep.net/search?p=find+Liu:2022chg}{{\ttfamily
  InSPIRE}}].

\end{thebibliography}
\end{document}